\documentclass[12pt, twoside]{article} 
\usepackage{amsmath} 
\usepackage{amsfonts} 
\usepackage{amsthm} 
\theoremstyle{definition} 
 
\newcommand{\field}[1]{\mathbb{#1}}

\newcommand{\HH}{\field{H}} 
\newcommand{\bo}[1]{\boldsymbol{#1}} 
\newcommand{\mf}[1]{\mathsf{#1}} 
\newcommand{\mc}[1]{\mathcal{#1}} 
\begin{document}
\title{Geometry of the group of nonzero quaternions} 
\author{Vladimir Trifonov \\ AMS \\ 
trifonov@member.ams.org} 
\date{}\maketitle 
\begin{abstract} 
It is shown that the group of nonzero quaternions carries a family of natural closed  Friedmann-Lema\^{\i}tre-Robertson-Walker metrics. \end{abstract} 

\section*{Introduction.} The quaternion algebra $\HH$ (e.~g. \cite{Wid02} and references therein) is one of the most important and well-studies objects in mathematics and physics (e.~g. \cite{Ad95} and references therein). It has a natural Hermitian form which induces a Euclidean metric on its additive vector space $S_{\HH}$. It also has a family of natural Minkowski metrics on $S_{\HH}$, induced by the structure constant tensor $\bo{\mf{H}}$ of the quaternion algebra \cite{Tri95}. We shall describe several natural structures on $\HH$ and $\mc{H} := \HH \setminus \{\bo{0}\}$.
\section{Natural coordinates.} The basis of unit quaternions $(\bo{1}$, 
$\bo{i}$, $\bo{j}$, $\bo{k})$ on $S_{\HH}$ is distinguished in the sense the it canonically 
generates the quaternion algebra $\HH$, i.~e., any four-dimensional real vector space $V$ containing a basis whose vectors acquire the multiplicative behavior of the unit  quaternions, becomes isomorphic to the quaternion algebra $\HH$. However, any other basis 
related to the canonical basis $(\bo{1}$, $\bo{i}$, $\bo{j}$, $\bo{k})$ on $S_{\HH}$ by a 
matrix 
\begin{displaymath} \begin{pmatrix} 1 & 0 \\ 0& \mf{B} \end{pmatrix}, \mf{B} \in 
SO(3). \end{displaymath} 
generates $\HH$ in exactly the same manner. Thus there is a class of canonical bases 
on $S_{\HH}$ whose members differ from one another by a rotation in the hyperplane of pure 
imaginary quaternions. Each canonical basis $(\bo{i}_{\beta})$ induces a ``perfect'' 
coordinate system $(w$, $x$, $y$, $z)$ on $S_{\HH}$, considered as a (linear) manifold, and 
therefore also on its submanifold $\mc{H}$ of nonzero quaternions: a quaternion $a = a^{\beta}\bo{i}_{\beta}$ 
is assigned coordinates $(w = a^0$, $x = a^1$, $y = a^2$, $z = a^3)$. This coordinate 
system covers both $S_{\HH}$ and $\mc{H}$ with a single patch. Notice that since $\bo{0} 
\notin \mc{H}$, at least one of the coordinates is always nonzero for any point $p \in 
\mc{H}$. For a real differentiable function $R \not\equiv 0$ there is a system of natural spherical coordinates  $(\eta$, $\chi$, $\theta$, $\phi)$ on $\mc{H}$, related to the canonical coordinates by 
\begin{multline*} 
w = R(\eta)\cos(\chi), \quad x = R(\eta)\sin(\chi)\sin(\theta)\cos(\phi), \\
y = R(\eta)\sin(\chi)\sin(\theta)\sin(\phi), \quad z = R(\eta)\sin(\chi)\cos(\theta) . 
\end{multline*}
\section{Natural basis fields.} Each canonical basis $(\bo{i}_{\beta})$ 
can be considered a basis on the vector space of the Lie algebra of $\mc{H}$, i.~e., 
the tangent space $T_{\bo{1}}\mc{H} \cong S_{\HH}$ to $\mc{H}$ at the point $(1$, $0$, 
$0$, $0)$, the identity of the group $\mc{H}$. There are several natural basis fields on 
$\mc{H}$ induced by each basis $(\bo{i}_{\beta})$. First of all, there are two coordinate 
basis fields, $(\partial_w$, $\partial_x$, $\partial_y$, $\partial_z)$ and the corresponding spherical coordinate basis $(\partial_{\eta}$, $\partial_{\chi}$, $\partial_{\theta}$, $\partial_{\phi})$. We also have a noncoordinate basis of left invariant vector fields on $\mc{H}$ (also denoted $(\bo{i}_{\beta})$), induced by a canonical basis. A left invariant vector field $\xi$ on $\mc{H}$, generated by a vector $\zeta \in S_{\HH}$, with components $(a, b, c, d)$ in the coordinate basis, associates to each point $p \in \mc{H}$ with coordinates $(w$, $x$, $y$, $z)$ a vector $\xi(p) \in T_p\mc{H}$ with the components $\xi^{\beta} = (p\zeta)^{\beta}$ in the coordinate basis: 
\begin{multline} \label{LVFIELDS} 
\xi^0 = wa - xb - yc - zd , \quad \xi^1 = wb + xa + yd - zc , \\ 
\xi^2 = wc - xd + ya + zb , \quad \xi^3 = wd + xc - yb + za . 
\end{multline} 
The system \eqref{LVFIELDS} contains sufficient information to compute transformation between the bases. 
The transformation between the canonical coordinate basis $(\partial_w$, $\partial_x$, $\partial_y$, $\partial_z)$ and left invariant basis $(\bo{i}_{\beta})$ is given by the matrix of the differential of left translations, easily computed from \eqref{LVFIELDS}: 
\begin{displaymath} \begin{pmatrix} w & -x & -y & -z \\ x & w & -z & y \\ y & z & w & -x \\ z & -y & x & w \end{pmatrix}.  \end{displaymath}
Each column of this matrix consists of components (in the coordinate basis) of the corresponding vector of $(\bo{i}_{\beta})$. Similarly, the transformation between the canonical coordinate basis $(\partial_w$, $\partial_x$, $\partial_y$, $\partial_z)$ and the spherical basis $(\partial_{\eta}$, $\partial_{\chi}$, $\partial_{\theta}$, $\partial_{\phi})$ is given by \small \begin{displaymath} 
\begin{pmatrix} \dot{R}\cos{\chi} & -R\sin{\chi} & 0 & 0 \\ 
\dot{R}\sin{\chi}\sin{\theta}\cos{phi} & R\cos{\chi}\sin{\theta}\cos{\phi} & R\sin{\chi}\cos{\theta}\cos{\phi} & -R\sin{\chi}\sin{\theta}\sin{\phi} \\ 
\dot{R}\sin{\chi}\sin{\theta}\sin{\phi} & R\cos{\chi}\sin{\theta}\sin{\phi} & R\sin{\chi}\cos{\theta}\sin{\phi} & R\sin{\chi}\sin{\theta}\cos{\phi} \\ 
\dot{R}\sin{\chi}\cos{\theta} & R\cos{\chi}\cos{\theta} & -R\sin{\chi}\sin{\theta} & 0 
\end{pmatrix},  \end{displaymath} \normalsize
where $\dot{R} := \frac{dR}{d{\eta}}$. Each column of this matrix consists of components (in the coordinate basis) of the corresponding vector of the spherical basis.  The transformation between the spherical and left invariant basis is given by 
\begin{displaymath} \begin{pmatrix} 
R/\dot{R} & 0 & 0 & 0\\ 
0 & \sin{\theta}\cos{\phi} & \sin{\theta}\sin{\phi} & \cos{\theta} \\ 
0 & \frac{\cos{\chi}\cos{\theta}\cos{\phi}+\sin{\chi}\sin{\phi}}{\sin{\chi}} & 
\frac{\cos{\chi}\cos{\theta}\sin{\phi}+\sin{\chi}\cos{\phi}}{\sin{\chi}} & 
\frac{\cos{\chi}\sin{\theta}}{\sin{\chi}}\\ 
0 & \frac{\sin{\chi}\cos{\theta}\cos{\phi}-\cos{\chi}\sin{\phi}}{\sin{\chi}\sin{\theta}} & 
\frac{\sin{\chi}\cos{\theta}\sin{\phi}+\cos{\chi}\cos{\phi}}{\sin{\chi}\sin{\theta}} & -1 
\end{pmatrix},\end{displaymath} 
where $\dot{R} \not\equiv 0$. 
\section{Natural extensions of the structure tensor.}
The structure tensor $\bo{\mf{H}}$ is a multilinear function of two vector arguments $a, 
b$ and a one-form argument $\tilde{\omega}$. Choosing a particular one-form $\tilde{\tau}$ (i. e. ether form) on $S_{\HH}$ makes the tensor $\bo{\mf{H}}(\tilde{\tau}; a, b)$ depend only on the vector arguments. Thus, if $\bo{\mf{H}}(\tilde{\tau}; a, b)$ is symmetric in $a$ and $b$, it is a (proper- or pseudo-)Euclidean metric on $S_{\HH}$. It was shown in \cite{Tri95} that every metric generated in this way is a Minkowski metric on $S_{\HH}$.
$\bo{\mf{H}}$ can be naturally extended in each of the above bases, by letting its components in a basis be constant and equal to the components at the identity of $\mc{H}$. However, it is easy to see that all left invariant bases generate the same extension, while each coordinate basis generates its own extension. Therefore the left invariant extension is ``more natural'' than the others, which is the case for an arbitrary tensor. The extension  (also denoted $\bo{\mf{H}}$) is given by the following four matrices
\begin{multline*} \mf{H}^0_{\alpha \beta} = 
\begin{pmatrix} 1&0&0&0\\0&-1&0&0\\0&0&-1&0\\0&0&0&-1 \end{pmatrix},\ 
\mf{H}^1_{\alpha \beta} = \begin{pmatrix} 0&1&0&0\\1&0&0&0\\0&0&0&1\\0&0&-
1&0 \end{pmatrix}, \\ \mf{H}^2_{\alpha \beta} = \begin{pmatrix} 
0&0&1&0\\0&0&0&-1\\1&0&0&0\\0&1&0&0 \end{pmatrix},\ \mf{H}^3_{\alpha 
\beta} = \begin{pmatrix} 0&0&0&1\\0&0&1&0\\0&-1&0&0\\1&0&0&0 
\end{pmatrix}. \end{multline*} 
These are the components of $\bo{\mf{H}}$ in a left invariant basis $(\bo{i}_{\beta})$ at any point $p \in \mc{H}$.
\section{Metric extensions.} A one-form $\tilde{\tau}$ with components $\tilde{\tau}_{\beta}$ in (the dual of) a canonical basis contracts with the structure tensor into the following tensor on $S_{\HH}$: 
\begin{displaymath} G_{\alpha \beta} = \begin{pmatrix} 
\tilde{\tau}_0& \tilde{\tau}_1& \tilde{\tau}_2& \tilde{\tau}_3\\ 
\tilde{\tau}_1&-\tilde{\tau}_0& \tilde{\tau}_3&-\tilde{\tau}_2\\ 
\tilde{\tau}_2&-\tilde{\tau}_3&-\tilde{\tau}_0& \tilde{\tau}_1\\ 
\tilde{\tau}_3& \tilde{\tau}_2&-\tilde{\tau}_1&-\tilde{\tau}_0 \end{pmatrix}. \end{displaymath} 
The only way to make this symmetric is to put $\tilde{\tau}_1=-\tilde{\tau}_1$, $\tilde{\tau}_2=-\tilde{\tau}_2$,  $\tilde{\tau}_3=-\tilde{\tau}_3$, which yields $\tilde{\tau}_1=\tilde{\tau}_2=\tilde{\tau}_3=0$. 
It is easy to see that in order to contract with $\bo{\mf{H}}$ into a metric on $\mc{H}$, a one-form \emph{field} $\tilde{\tau}(p)$ must have the components 
$(\tau(p)$, $0$, $0$, $0)$ in the left invariant basis $(\bo{i}_{\beta})$ at every point $p \in \mc{H}$.  In the spherical basis $\tilde{\tau}$ has components $((R/\dot{R})\tau(p)$, $0$, $0$, $0)$ (where $\dot{R} := dR/d{\eta}$), which means that $\tau$ must depend only on $\eta$. Then its contraction with $\bo{\mf{H}}$ produces a natural metric whose components in the spherical basis are 
\begin{displaymath} g_{\alpha \beta} = \begin{pmatrix} 
\tau(\eta)(\frac{\dot{R}}{R})^2&0&0&0\\0&-\tau(\eta)&0&0\\ 
0&0&-\tau(\eta){\sin^2(\chi)} &0\\ 
0&0&0&-\tau(\eta){\sin^2(\chi)} {\sin^2(\theta)} \end{pmatrix}. \end{displaymath} 
If $\tau(\eta) > 0$, we take $R(\eta)$ such that $\tau(\eta)(\frac{\dot{R}}{R})^2 = 
1$, which yields \begin{equation} \label{+R} R(\eta) = 
exp{\int\frac{d\eta}{\genfrac{}{}{0pt}{3}{+}{-}\sqrt{\tau(\eta)}}} ; \quad 
\tau(\eta) > 0 . \end{equation} 
In other words, with $R(\eta)$ satisfying \eqref{+R}, the metric is 
closed Friedmann-Lema\^{\i}tre-Robertson-Walker: \begin{displaymath} 
g_{\alpha \beta} = \begin{pmatrix} 
1&0&0&0\\ 0&-\mf{a}^2&0&0\\ 0&0&-\mf{a}^2{\sin^2(\chi)} &0\\ 
0&0&0&-\mf{a}^2{\sin^2(\chi)} {\sin^2(\theta)} \end{pmatrix}, \end{displaymath} 
where the ``expansion factor'' $\mf{a}(\eta) := {\sqrt{\tau(\eta)}}$. \par 
If $\tau(\eta) < 0$, we take $R(\eta)$ such that $\tau(\eta)(\frac{\dot{R}}{R})^2 = 
-1$, and we obtain \begin{equation} \label{-R} R(\eta) = 
exp{\int\frac{d\eta}{\genfrac{}{}{0pt}{3}{+}{-}\sqrt{-\tau(\eta)}}} ; \quad 
\tau(\eta) < 0 . \end{equation} 
Hence with $R(\eta)$ satisfying \eqref{-R}, the metric is also closed FLRW: 
\begin{displaymath} g_{\alpha \beta} = \begin{pmatrix} 
-1&0&0&0\\ 0&\mf{a}^2&0&0\\ 0&0&\mf{a}^2{\sin^2(\chi)} &0\\ 
0&0&0&\mf{a}^2{\sin^2(\chi)} {\sin^2(\theta)} \end{pmatrix}, \end{displaymath} 
where the ``expansion factor'' $\mf{a}(\eta) := {\sqrt{-\tau(\eta)}}$. 
\par Thus the natural geometry of the group of nonzero quaternions $\mc{H}$ is defined by a family of closed  Friedmann-Lema\^{\i}tre-Robertson-Walker metrics. 
\section{Integral curves of LI vector fields on $\mc{H}$.} 
The system \eqref{LVFIELDS} can be presented as a system of linear differential equations 
\begin{multline} \label{DIIFLFIELDS}
\dot{w} = aw(t) - bx(t) - cy(t) - dz(t),\\ 
\dot{x} = bw(t) + ax(t) + dy(t) - cz(t),\\ 
\dot{y} = cw(t) - dx(t) + ay(t) + bz(t),\\ 
\dot{z} = dw(t) + cx(t) - by(t) + az(t), 
\end{multline} 
Given a point $\bar{p} \in \mc{H}$ with coordinates $(\bar{w}, \bar{x}, \bar{y}, \bar{z})$ 
a solution of \eqref{DIIFLFIELDS} is (the parametric form of) an integral curve, through $\bar{p}$, of the left invariant vector field on $\mc{H}$, generated by $\zeta \in \HH$: 
\begin{multline*} w(t) = e^{at}(W\sin{\omega{t}} + \bar{w}\cos{\omega{t}}), \\ 
x(t) = e^{at}(X\sin{\omega{t}} + \bar{x}\cos{\omega{t}}), \\ 
y(t) = e^{at}(Y\sin{\omega{t}} + \bar{y}\cos{\omega{t}}), \\  
z(t) = e^{at}(Z\sin{\omega{t}} + \bar{z}\cos{\omega{t}}), \end{multline*} 
where $\omega := \sqrt{b^2 + c^2 + d^2}$, and
\begin{multline*}  W := \frac{-\bar{x}b - \bar{y}c - \bar{z}d}{\omega}, \quad X := \frac{\bar{w}b - \bar{z}c + \bar{y}d}{\omega}, \\
Y := \frac{\bar{z}b + \bar{w}c - \bar{x}d}{\omega},  \quad Z := \frac{-\bar{y}b + \bar{x}c + \bar{w}d}{\omega}. \end{multline*}

\end{document}